\documentstyle[11pt,aaspp]{article}

\input psfig

\newcommand{\eg}{{\it e.g.}\ }

\newcommand{\etal}{{\it et~al.}\ }

\newcommand{\ale}{\ \raisebox{-.3ex}{$\stackrel{<}{\scriptstyle \sim}$}\ }
\newcommand{\age}{\ \raisebox{-.3ex}{$\stackrel{>}{\scriptstyle \sim}$}\ }

\begin{document}

\title{ACCRETION DISKS IN INTERACTING BINARIES:\\ 
		  SIMULATIONS OF THE STREAM-DISK IMPACT}
\author{P.J. Armitage\altaffilmark{1}}

\author{M. Livio\altaffilmark{2}}

\altaffiltext{1}{Institute of Astronomy, Madingley Road, 
		 Cambridge, CB3 0HA, UK}
\altaffiltext{2}{Space Telescope Science Institute, 3700 San Martin Drive, 
       Baltimore, MD 21218, USA.}

\begin{abstract}
We investigate the impact between the gas stream from the 
inner Lagrangian point and the accretion disk
in interacting binaries, using three dimensional Smooth Particle Hydrodynamics
simulations. We find that a significant fraction of the stream material can
ricochet off the disk edge and overflow towards smaller radii, and that this
generates pronounced non-axisymmetric structure in the absorption column
towards the central object. We discuss the implications of our results
for observations and time-dependent models of low-mass X-ray binaries, 
cataclysmic variables and supersoft X-ray sources.

\bigskip

\centerline{\it ApJ accepted}

\end{abstract}

\newpage

\section{INTRODUCTION}
Accretion disks in interacting binaries are unlikely to have
simple axisymmetric structures. In these systems, the accretion
disk is fed with material from a companion star via an
accretion stream that originates at the inner Lagrange point.
In this picture, both the gravitational influence of the companion 
and the impact of the stream with the disk provide mechanisms that
may distort the shape and vertical structure of the accretion
disk. Modeling of these deviations from axial symmetry requires the use of
two and three dimensional hydrodynamic models, and is necessary
in order to interpret a variety of observations of cataclysmic 
variables (CVs), low-mass X-ray binaries (LMXBs) and 
supersoft x-ray sources (SSS).

The X-ray light curves of nearly edge-on LMXBs provide one
observational clue to the existence of non-axisymmetric
disk structures. The eclipses are typically broad, with 
a gradual ingress that commences well before the phase of first 
contact (\eg White \etal 1994). In some sources (``dipping'' sources), most 
notably X1822-371, there is an actual dip in the flux at an
orbital phase near 0.8. This absorption has been interpreted
as arising from a bulge in the height of the disk rim at
that phase (White \& Holt 1982; Mason 1989 and references therein), 
which is suggestively similar
to the location at which the accretion stream would impact
the disk edge. X-ray observations of GK Persei (Hellier
\& Livio 1994) and EUVE observations of CVs (Warren, Sirk \& 
Vallerga 1995; Long \etal 1996) provide further evidence for observational
signatures of the accretion stream and its interaction
with the disk.

Two dimensional calculations of accretion disk structure 
have primarily been used to investigate the formation of the disk 
and its response to the tidal field of the companion
(Lin \& Pringle 1976; Whitehurst 1988a,b; Lubow 1991;
Murray 1996). Using a variety of particle-based numerical
schemes, these studies show that for systems with a low
mass ratio of secondary to primary ($q \ale 0.25$), the 
accretion disk is tidally
unstable, leading to the formation of an eccentric
precessing disk. Proper consideration of the effect
of the accretion stream on the disk (which acts to add
low specific angular momentum material at the outer edge
and so tends to reduce the disk radius) shows that this 
modifies the phenomenon but does not alter the general
behaviour (Murray 1996).

Hirose, Osaki \& Minishige (1991) carried out simplified three dimensional
calculations in an attempt to examine whether variations
in disk scale height caused by the stream impact could explain
the dipping X-ray sources. They found thickenings of the 
disk at around the phases expected, if these were the causes
of modulation in the X-ray light curves. However, the difficulties
of modeling the entire disk for many dynamical times meant that
the resolution of the stream-disk impact region in this calculation 
was poor, and the authors commented that an improved treatment
of the hydrodynamics beyond their ``sticky particle'' scheme was
desirable. Much better resolution of the impact region was
achieved in calculations by Livio, Soker \& Dgani (1986) and
Dgani, Livio \& Soker (1989), but this was achieved at the expense 
of considering only a small region surrounding the point
where the stream met the disk. Thus, while these calculations
demonstrated clearly the flow pattern in the collision region,
there remained the possibility that the hydrostatic vertical structure
assumed for the disk gas might not be realistic, if the response of 
the entire disk to the stream was included. With this caveat,
however, the calculations did demonstrate that stream material could
overflow the disk rim, as suggested by the work of Lubow \& Shu
(1975, 1976), Lubow (1989) and Frank, King \& Lasota (1987).

In the current investigation, we extend these studies by 
performing three dimensional Smooth Particle Hydrodynamics (SPH)
simulations of the response of the disk to the accretion stream.
We focus on the dynamical effects of the stream-disk collision -- for
which the SPH method is well-suited -- and aim for the highest
practicable resolution of the interaction region. We do not 
attempt to model the thermal or long-term viscous evolution
of the disk, for which purpose other numerical techniques
would be preferable. By ignoring these aspects we are able to 
improve the resolution of the stream-disk impact region when compared,
for example, to
prior 3D simulations by Lanzafame, Belvedere \& Molteni (1994a,
1994b), who were interested primarily in modeling outburst
behaviour over longer timescales than those considered here.

The layout of this paper is as follows. In Section 2 we describe
the computational methods and binary system parameters that we
adopt for the simulations. Section 3 presents and interprets
the results of the calculations, while Section 4 summarizes 
our conclusions and suggests further observational tests
of the models.

\section{COMPUTATIONAL APPROACH}

\subsection{Smooth Particle Hydrodynamics}
The simulations described in this paper were performed using
a three dimensional SPH code minimally modified from that
described by Benz (1990). SPH is a fully lagrangian, particle-based 
hydrodynamics method, that has been applied
to a wide variety of problems in astrophysics (see, \eg Monaghan
1992), and has been shown to give closely comparable results
to hydrodynamics codes based on alternative numerical techniques
(\eg PPM---Davies \etal 1993).

For the simulation of accretion disks, SPH has a number of
advantages. A Keplerian disk exhibits a rapid variation of
dynamical timescale with radius, which is computationally
expensive if the problem demands that a significant range
of radii be modeled. This difficulty can be partially alleviated in
an SPH calculation by assigning each particle its own 
timestep satisfying the Courant stability criterion, so
that the bulk of the disk material at relatively large
radii can be integrated on a longer timestep. SPH 
has further merits in facilitating easy tracing of the
flow, and in not wasting resources following the void
regions outside and above the accretion disk.

Disadvantages of this method include the variable resolution of an SPH
calculation, which is worst in the low density parts
of the flow where the particle density is small. As these
regions---well above the mid-plane of the disk---are
precisely those that are most important for determining
the X-ray and EUV absorption, this is a potentially
serious problem. We mitigate its effect by exploiting
the freedom that SPH allows in setting particle masses,
so that the stream particles (which on collision with the
disk will form the majority of the material high above the
disk plane) have lower masses that those in the disk. More
problematic is the fact that our code neglects the effects
of radiation transport and cooling of the gas, so that the
only control over the thermal properties arises via the choice
of the equation of state. We adopt an isothermal equation of
state, which corresponds to assuming that the energy generated 
from viscous and shock heating is radiated away instantaneously.
For a standard $\alpha$ accretion disk (Shakura \& Sunyaev 1973),
the thermal timescale,
$t_{\rm th} \sim \alpha^{-1} \Omega^{-1}$, is comparable to
the dynamical timescale, $t_{\rm dyn} \sim \Omega^{-1}$, if
the Shakura-Sunyaev viscosity parameter $\alpha$ is of
order unity (Pringle 1981). Arguments based on the timescales
involved in the outbursts in these systems suggest that for CVs and
LMXBs $\alpha$ is typically in the range 0.1--1, so that
the isothermal equation of state provides a reasonable
approximation to the much more complicated physics that is
actually involved. Caution is nevertheless required in
interpreting our results close to where shocks---and
hence strong and rapid heating of the gas---occur in
the simulations.

\subsection{System Parameters}

Having ignored both thermal and viscous effects, the important
system parameters that remain are the mass ratio $q = M_2/M_1$, the
disk mass $M_{\rm disk}$, and the accretion rate $\dot M$. 
For low mass ratios $q \ale 0.25$, prior calculations
show the formation of an eccentric, tidally unstable disk,
which precesses in the rotating frame of the binary (Whitehurst 1988a,b; 
Lubow 1991). Such mass ratios are likely to be relevant both to
subclasses of dwarf novae (the SU UMa systems), and to many
LMXBs with low-mass (a few tenths $M_\odot$) companions. 
Conversely, higher mass ratios lead only to a modestly distorted
disk which remains fixed in the Roche frame of the binary.

The disk mass and accretion rate are expected to be significant
mainly via the combination $f = \dot{M} P / M_{\rm disk}$, where
$P$ is the period of the binary. $f$ is then the fraction
of the disk mass that is replenished per orbit, and it measures 
how strong a perturbation the stream can exert on the disk.
To estimate sensible values for $f$, we assume a Shakura-Sunyaev 
disk in which the
viscosity is given by $\nu = \alpha c_{\rm s} H$, where $c_{\rm s}$
is the sound speed and $H \approx c_{\rm s} / \Omega$ is the disk 
scale height. For such a disk,
the surface density $\Sigma$ in the steady state is given by,
\begin{equation}
 \nu \Sigma = {\dot{M} \over {3 \pi}},
\label{1}
\end{equation}
for radii much larger than the inner disk radius. Taking 
$M_{\rm disk} \sim \pi R^2 \Sigma$, we then obtain an 
estimate for f,
\begin{equation}
 f = { {\dot{M} P} \over M_{\rm disk} } \sim 6 \pi \alpha \left( c_{\rm s}
 \over v_{\rm K} \right)^2,
\label{2}
\end{equation}
where $v_{\rm K}$ is the Keplerian velocity in the disk and 
all quantities are evaluated at the disk edge. For a thin
disk $c_{\rm s} / v_{\rm K} \ll 1$ ($\approx 0.03$ for our
simulations), so that $f$ is small, typically of order $10^{-2}$,
even for $\alpha$ of order unity. This is also consistent with the
timescales of outbursts (in which mass accumulated in the disk 
over an extended time rapidly drains onto the central star),
which generally last $\sim 10^2 P$.

In previous calculations the most dramatic effects of the
companion on the accretion disk were found to occur for
low mass ratios. We therefore adopt $q = 0.2$, as appropriate
for a $0.3 \ M_\odot$ companion to a $1.4 \ M_\odot$ primary,
and $f = 0.025$. For the particular value of $M_{\rm disk}$
in the calculations, this corresponds to a mass transfer
rate of $\dot{M} = 2 \times 10^{17} \ {\rm gs^{-1}}$, or 
$3 \times 10^{-9} \ M_\odot \ {\rm yr^{-1}}$ (run LMXB2).
To assist in disentangling the effects of the Roche potential
from those caused solely by the impact of the stream, a second
simulation was run with a very low mass secondary ($q$ effectively 
zero), and otherwise similar values of disk mass and accretion rate
(run LMXB1).

\subsection{Disk and Stream Set Up}

As we have already remarked, simulating the innermost regions of the
disk is computationally prohibitive. Fortunately it is also 
unnecessary, since on a dynamical timescale the stream is unable
to influence the disk interior to the stream circularization
radius. We therefore set up the disk as an annulus with an inner
edge at approximately the circularization radius and an outer
edge at the tidal truncation radius as given by Papaloizou \& 
Pringle (1977) and Paczy\'nski (1977). To prevent a small
fraction of particles from bringing the simulation to a halt by
moving too far inward under the action of viscosity, we add
angular momentum to any disk particle straying inside a radius
$R_{\rm bound}$ from the primary. For the runs described here,
$R_{\rm bound}$ is set at half the inner radius of the annulus.
Over the relatively short timescale of our runs, we observe
no effect of this inner boundary condition on the disk annulus
farther out.

Although the effects of {\em physical} viscosity are unimportant
for these calculations, it is important that the artificial
viscosity in the code, parameterized by $\alpha_{\rm SPH}$ and 
$\beta_{\rm SPH}$ (Benz 1990), is sufficient to prevent particle 
interpenetration in the shocks that occur when the stream strikes the 
disk edge. An isothermal equation of state provides the least resistance
to particle interpenetration (as there is no shock heating to
generate additional pressure gradients), and so we adopt the
rather high values $\alpha_{\rm SPH} = 2$, $\beta_{\rm SPH} = 4$.
Previous work (Bate 1995) has shown that this should be
adequate to stop particles streaming past each other, and we
do not see any such behaviour in our runs. The internal SPH
viscosities can be converted into a physical Shakura-Sunyaev
$\alpha$ parameter via the formula (\eg Bate 1996),
\begin{equation}
 \alpha \approx \left(\alpha_{\rm SPH} \over 20 \right) 
 \left( h \over H \right),
\label{3}
\end{equation}
where $h$ is the particle smoothing length in the simulation (3D).
The $\beta$-viscosity will also have some effect, but
for a disk that is resolved in the vertical direction, the
contribution to the kinematic viscosity from the second-order
$\beta_{\rm SPH}$ term will be smaller than that from 
$\alpha_{\rm SPH}$ (Bate 1996).
Using this expression it can be seen that for simulations that
are resolved vertically ($h < H$, as in our calculations), typical
SPH viscosity parameters lead to values of $\alpha$ that are
smaller than those believed to pertain to CV and LMXB
accretion disks, and there should be no problems arising
from unreasonably large artificial viscosity in the simulations.

\begin{figure}[t]
 \hspace{0.9truein}
 \psfig{figure=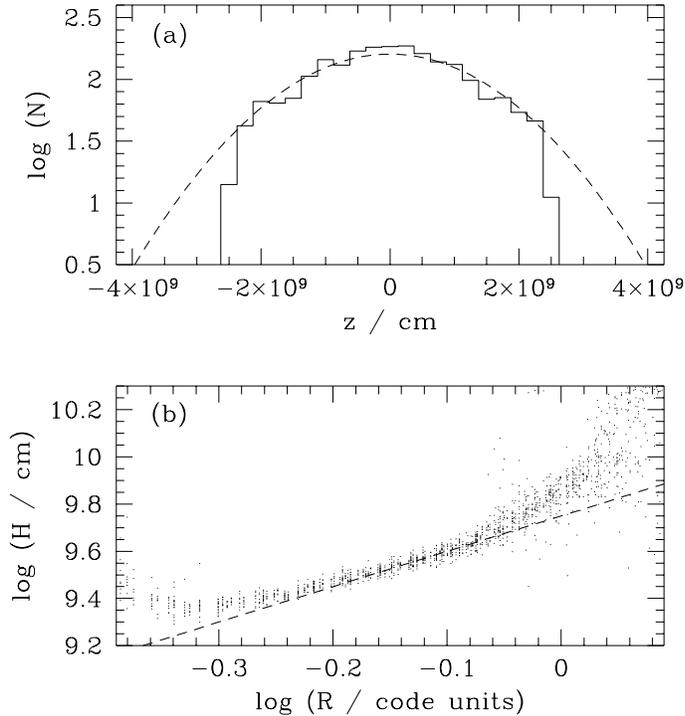,width=4.0truein,height=4.0truein}
 \caption{(a) Distribution of particles in $z$ for a thin annulus in
a quiescent (no stream) disk. The dashed line shows a gaussian
distribution with width $\Delta z = 2 \times 10^{9}$ cm. (b) The
disk scale height evaluated in cells for the same quiescent disk,
plotted as a function of radius. The dashed line represents 
$H \propto R^{3/2}$.} 
 \label{fig1}
\end{figure}

The disk gas is modeled using 30000 equal mass particles that
are initially distributed on a close-packed lattice within
$2.5^{\circ}$ of the disk plane, and which have Keplerian
velocities about the primary. The sound speed is set such
that $c_{\rm s} / v_{\rm K} = 0.03$ at the outer edge of the disk---this
gives a somewhat thicker disk than is usually assumed, but has the
advantage that better resolution is attained in the vertical 
direction. The disk reaches a stable hydrostatic equilibrium
structure quickly (in less than one orbit), after which the distribution
of particles with $z$ (perpendicular to the disk) for a narrow 
annulus of disk is as shown in
Fig. 1(a). The particles follow a gaussian distribution, as expected
for an isothermal accretion disk, and track the density profile
well, up to the height where the particle number density becomes
very low.
Calculating the disk scale height 
(defined as $H = \Sigma / \rho_{z=0}$, where $\Sigma$ and $\rho_{z=0}$ are 
evaluated from the SPH calculation) 
as a function of radius yields
the results shown in Fig. 1(b).
For an isothermal disk,
$H \propto c_{\rm s} / \Omega \propto R^{3/2}$, and this
scaling is shown in the Figure as the dashed line. Over
the range ($-0.3 < \log (R) < 0$) where the disk is
resolved vertically this relation is found to be obeyed,
with discrepancies at the inner and outer edges where 
the particle density is dropping off rapidly.

For run LMXB2 the presence of the companion means that
the disk is not being set up with velocities that are
consistent with disk orbits in the binary potential. 
To allow time for these initial conditions
to be smoothed out, the disk is first evolved for
$\sim 10$ orbits of the binary before the stream flow
is switched on. During this relaxation period the
disk develops a stable but distorted shape, with a
smooth variation of surface density. For the purposes
of the present work this represents
well enough the `background' disk structure onto
which the stream flows, though if,
for example, one wished to follow the disk precession 
in a 3D calculation, a more elaborate disk set up would
certainly be desirable. 

For the initial conditions of the stream, we utilise the results
of Lubow \& Shu (1975, 1976), see also review by Livio (1994).
These show that the width of the stream, $W$, in both
the azimuthal and vertical directions, is given approximately by,
\begin{equation}
 W \sim {c_{\rm s} \over \Omega_{\rm binary}},
\label{width}
\end{equation}
where $c_{\rm s}$ is here the sound speed at the surface of the
mass-losing star and $\Omega_{\rm binary}$ is the binary
angular velocity. This is essentially the same relation as
for the disk scale height, and implies that the stream is
much thicker than the pressure scale height of the mass-donating
star $H_*$ (indeed, $W \sim \sqrt{H_* R_*}$).	

Numerically, the stream is modeled as a tube of particles stretching
outward from the inner Lagrange point, ${\rm L}_1$, with 
radial velocities
set to be $c_{\rm s}$ towards the primary. When a particle
passes inward of ${\rm L}_1$ (towards the primary) it is added to the 
simulation
and evolved under the influence of gravity and all hydrodynamic
forces, while particles outside ${\rm L}_1$ are simply rotated to
follow the binary orbit. Initially the stream particles
are strongly compressed in the radial direction (by a factor
$\sim 10$). This ensures that when the stream reaches the disk 
edge the separations are comparable in $R$ and in $z$, as
required to model the impact hydrodynamics. We assume that
the stream leaves ${\rm L}_1$ in hydrostatic equilibrium, 
and set the vertical density profile to be a gaussian
with width $W$. This
is truncated at 1 scale height, where the scale height is
estimated by extrapolating the numerical results obtained
from the disk (Fig.~1), and allowing, where necessary, for the gravity
of the secondary. The compression required in $R$ near 
${\rm L}_1$ means that the stream there is not in numerical
hydrostatic equilibrium, but this is of no consequence, since
the stream flow toward the disk is ballistic. The simulations
follow the stream-disk interaction for around 2-3 orbits of
the outer disk, requiring 20000--30000 particles to 
achieve adequate resolution of the accretion stream.

\section{RESULTS}

\subsection{Stream Overflow}

\begin{figure}[t]
 \caption{Particle distribution in the $x-y$ and $x-z$ planes for
simulation LMXB2 (mass ratio of secondary to primary
$q=0.2$). This figure is omitted due to size, and is available from 
{\it http://www.ast.cam.ac.uk/\~ \,parm/lmxb.html}} 
 \label{fig2}
\end{figure}

Fig.~2 shows a snapshot of particle positions during the
latter stages of run LMXB2. For both simulations, a 
quasi-steady state appearance in the Roche frame is rapidly 
attained, so that although the disk mass is slowly increasing
throughout, successive snapshots look very similar. Where the
stream reaches the disk, the Figure shows that the oblique
collision both shears the stream and compresses the disk
edge. Some stream material is transferring its
momentum to the disk gas and becoming buried in the
disk rim. This is similar to the behaviour seen in
previous calculations (\eg Livio \etal 1986; Rozcyczka \&
Schwarzberg-Czerny 1987). However,
from the $x-z$ projection it is also clear that some
stream gas is overflowing the disk rim as a result
of the collision. This material is being thrown
to larger distances above the midplane than the original stream,
and is many scale heights above the disk itself. 

\begin{figure}[t]
 \hspace{0.9truein}
 \psfig{figure=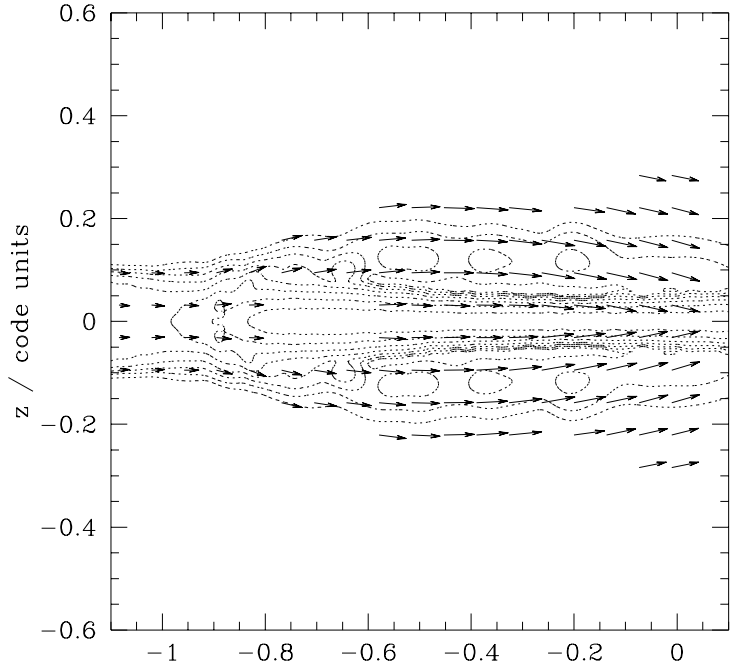,width=4.0truein,height=4.0truein}
 \caption{Logarithmic density contours (dashed lines) in a vertical
plane aligned with the direction of stream flow at the location
where the stream strikes the disk. Vectors represent the velocity
in this plane of stream material. The $z$-component of the velocity
vectors has been enhanced by a factor of two for clarity.} 
 \label{fig3}
\end{figure}

To clarify the origin of this overflowing gas, we
plot in Fig. 3 density contours in a vertical plane,
aligned along the direction of the accretion stream
flow where it reaches the disk edge. We also show
velocity vectors in this plane for the stream gas,
calculated by simple averaging of particle velocities in 
spheres of radius $2h$. Although the disk in general
flares towards large radius ($H \propto R^{3/2}$), it
can be seen that in the outermost regions of the disk
where the surface density is dropping, the density
contours form a wedge-like structure. The stream
strikes this supersonically, and the outer parts
ricochet off to greater distances from the disk 
midplane. This behavior has been predicted by the analytic Sedov Solution
obtained by Livio \etal (1986).  The vertical velocities away from the
$z=0$ plane that are achieved in this manner, are
typically several times $c_{\rm s}$, so this gas
reaches a maximum $|z|$ that is many times the
disk scale height. Further downstream the overflowing gas
falls back to the disk, achieving higher velocities
than on impact, because now it is closer to the primary 
where the vertical component of gravity is greater.
While the central core
of the stream tends to blunt the wedge, it is continually
restored by fresh disk material that possesses a higher
angular velocity than the orbital one.

\subsection{Density Slices from the Simulations}

\begin{figure}[t]
 \psfig{figure=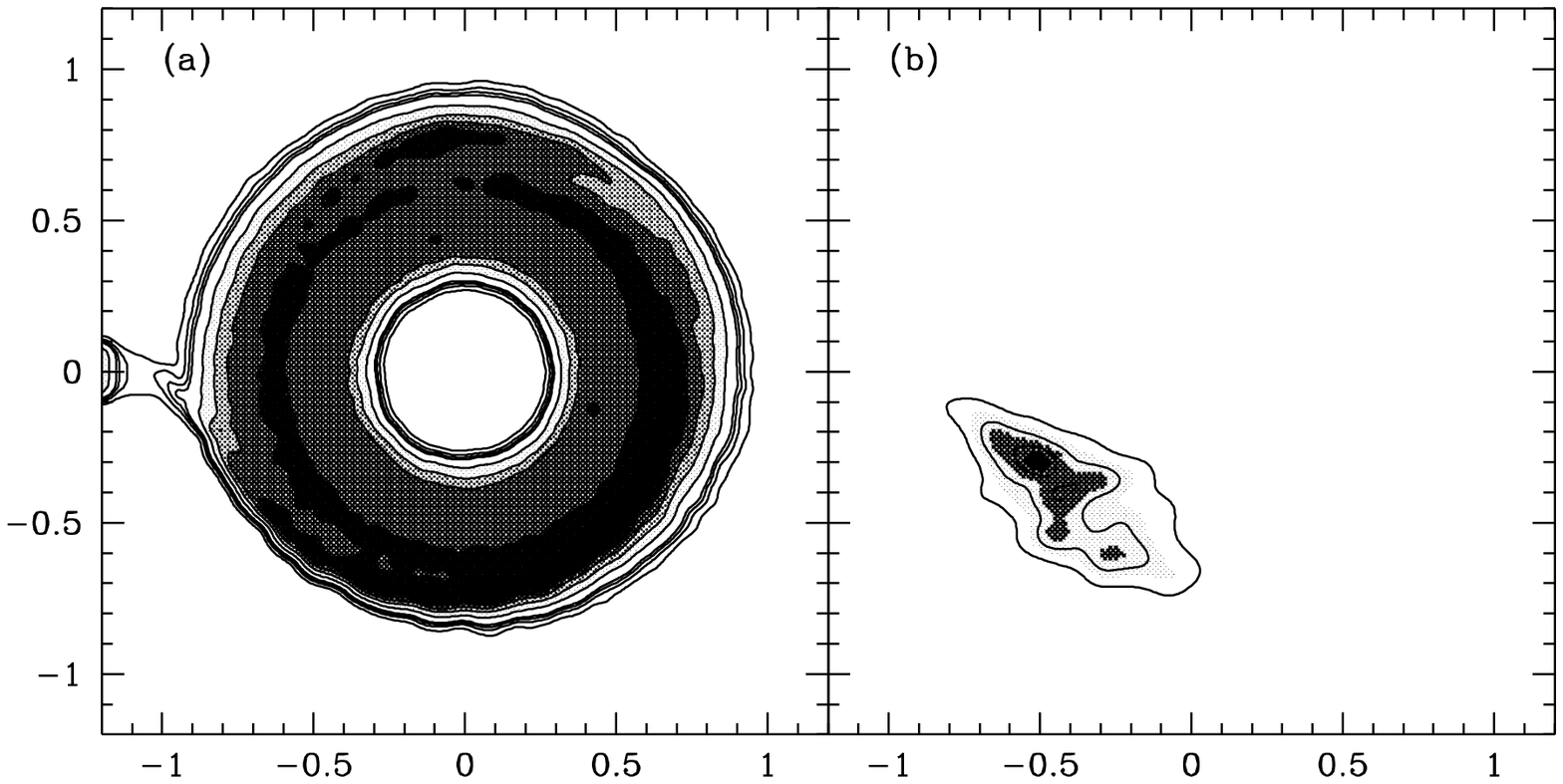,width=5.8truein,height=5.8truein}
 \vspace{-2.7truein}
 \caption{Density from simulation LMXB1 (secondary of zero mass) 
in the $x-y$ plane at (a) $z=0$, and (b) $z=0.15$.} 
 \label{fig4}
\end{figure}

Figs. 4 and 5 show the density calculated using the
SPH smoothing algorithm for both simulations. We plot
the midplane density, panel (a), and the density at 
a few scale heights above the midplane, panel (b).
In the latter plot, we have co-added 4 independent
time-slices in order to reduce the Poisson noise
due to limited particle number, though the behavior
we describe is clearly seen also in individual snapshots.
For both simulations the compression of the disk
rim due to the momentum of the stream impact is
evident (the oscillations in radius of the disk,
downstream of the impact, are probably due to
remaining anisotropies in particle distribution
on impact, and are not physical). However, except
for this, the midplane density plots show little
structure beyond what would be expected given
the initial set up of these disks.

At high $|z|$, conversely, strong signatures of the stream
impact are visible. For LMXB1 a single region of relatively
high density is obtained, beginning at the point where
the stream reaches the disk and persisting until around
phase 0.75 (here phases are measured with respect to the
phase at which an eclipse would occur, and increase clockwise
in the plots). This peak lies well inside the outer edge
of the disk, and arises from the overflowing stream
material discussed earlier. Although the resolution at
this $z$ is limited by relatively small particle numbers, 
it appears that the overflowing gas does {\em not}
form a well-collimated continuation of the accretion
stream. Rather the impact leads to a broad fan of
material ejected to high $|z|$ as a consequence of the
collision.

\begin{figure}[t]
 \psfig{figure=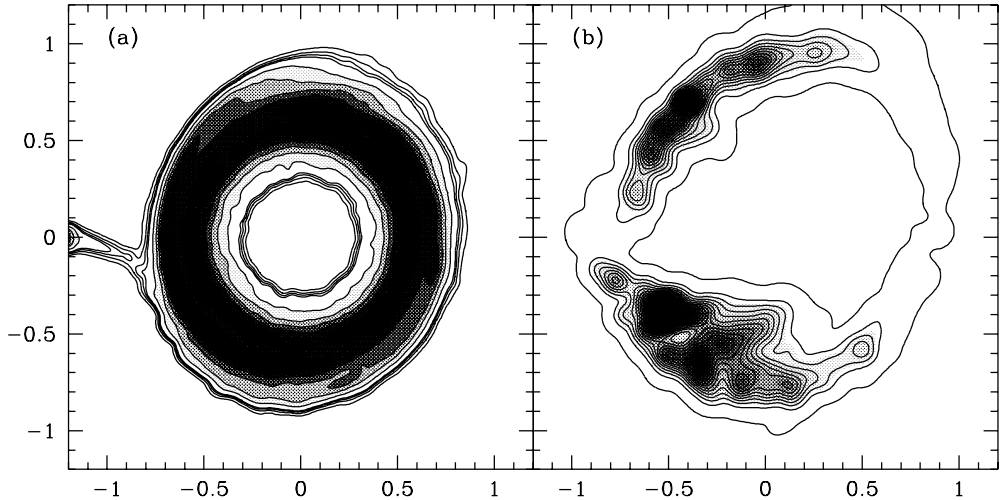,width=5.8truein,height=5.8truein}
 \vspace{-2.7truein}
 \caption{Density from simulation LMXB2 (full Roche potential, $q = 0.2$) 
in the $x-y$ plane at (a) $z=0$, and (b) $z=0.12$.} 
 \label{fig5}
\end{figure}

For the simulation with the full Roche potential, a 
similar structure generated by the stream-disk interaction
is seen in the density plots. In this instance however, a
second broad and comparatively high density region is seen at 
phase 0.2, which originates from the eccentricity of the
accretion disk in this simulation. The disk is flared,
and so for a given high $z$ slice, the greatest density contribution
is expected to come from material at the largest radius.
This leads to a peak roughly in the direction along which
the disk is most extended, though from the simulation the
peak is seen at a somewhat earlier phase, probably as a
consequence of the disk gas not reaching hydrostatic
equilibrium and hence lagging in adjusting to the
changing vertical gravity. We note that the magnitude
of the second peak is probably overestimated by the
simulation, as the SPH particles at the outer edge of
the disk have no neighbours at greater radii and
so adopt larger smoothing lengths to maintain a roughly
constant number of neighbours. These larger smoothing lengths
will enhance the density contribution at high $|z|$. When
this numerical effect is borne in mind, both simulations
demonstrate that material directly thrown above the 
midplane by the stream-disk impact is likely to dominate
the absorption column at high $|z|$. 

\subsection{Column Densities Towards the Primary}

In order to quantify the observable effects of the material 
thrown out of the plane
we have calculated the column density $\sigma$
along various lines of sight toward the primary from
our simulations. We calculate,
\begin{equation}
 \sigma (\phi,\theta) = \int \rho {\rm d}l
\label{4}
\end{equation}
from integration of the SPH density estimator along lines of
sight toward the primary at phase $\phi$ and at angle
$\theta$ above the disk midplane. $\phi$ values are
chosen so that neighbouring lines of sight are calculated
from different sets of particles and hence independent.
As the innermost particles have overly large smoothing lengths
for the reasons already discussed, these are removed before
column densities are calculated. This cut affects only
$\sim$ 1\% of the particles.

\begin{figure}[t]
 \hspace{0.5truein}
 \psfig{figure=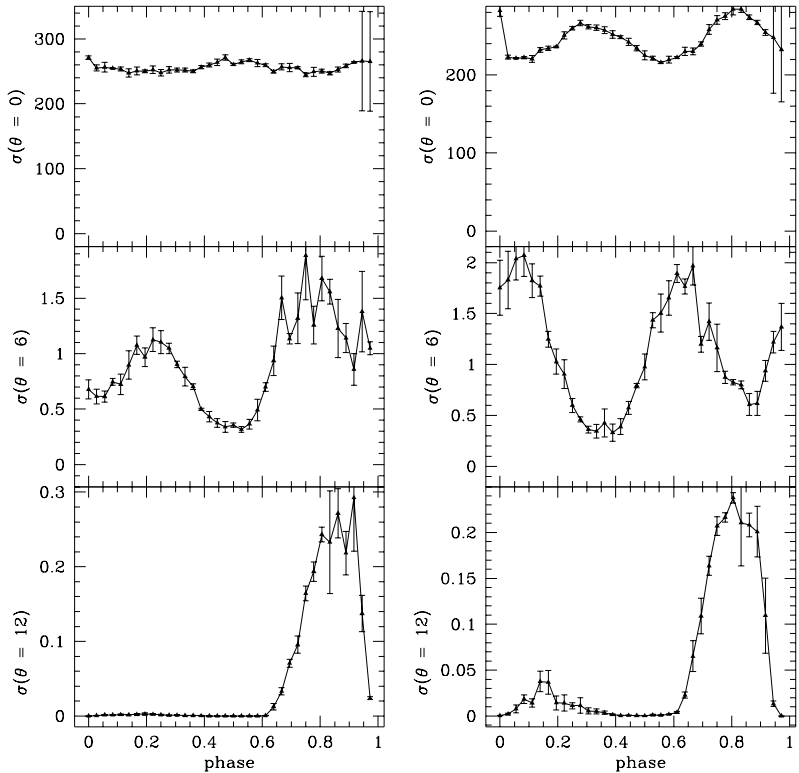,width=4.5truein,height=4.5truein}
 \caption{Column density $\sigma$ towards the primary as a function
of phase for varying angles of elevation above the disk
plane $\theta = 0, 6, 12$ degrees. Left panels are for simulation
LMXB1, right panels for LMXB2. Units are gcm$^{-2}$. Error bars
are calculated from 4 independent time-slices of the simulations.} 
 \label{fig6}
\end{figure}

Fig.~6 shows the results for both simulations, for angles
of elevation $\theta = 0, 6, 12^{\circ}$. For comparison,
the dipping source X1822-371 is believed to be viewed
at an inclination angle of $82^{\rm o}$--$87^{\rm o}$ (Mason 1989).
For LMXB1 there is no variation evident above the noise at
$\theta = 0^{\circ}$. By 
$\theta = 6^{\circ}$ the overflowing stream material, seen
in Fig. 4 at high $|z|$, is producing a significant contribution
to the column density, leading to a peak in $\sigma$ at phase
$\phi \sim 0.8$. There is marginal evidence for a broad
feature also at $\phi \sim 0.2$, caused by the compressed 
disk rim expanding outwards downstream of the stream impact point.
However, at the highest angle of elevation, $\theta = 12^{\circ}$,
the overflowing stream gas provides the {\em only} source of absorbing
column, and there is a single strong peak centred on $\phi = 0.85$
and extending from phase 0.7 through to eclipse. For an opacity
generated solely by Thomson scattering a column density of
order unity is required to produce an optical depth of 1, and
hence for this accretion rate ($3 \times 10^{-9} \ M_\odot \ {\rm yr^{-1}}$)
the overflowing gas would be sufficient to create significant
absorption even at angles to the midplane $\theta \age 10^{\circ}$.

For run LMXB2 the situation is more complicated. At $\theta = 12^{\circ}$
there is a single strong peak in column density at phase 0.8, which as before
is associated with the stream-disk interaction. There may be 
a hint of a very small
peak at $\phi \sim$ 0.1--0.2.  However in this
simulation there is significant variation in $\sigma$ with phase
even at $\theta = 0^{\circ}$, as a result of the marked distortion
of the disk shape under the gravitational influence of the companion.
Of course viewed directly edge-on, the optical depth to the central
object is enormous, and so the variation in this regime would
not be observable. These variations suggest, however, that at intermediate
viewing angles both effects should be important, and this is 
seen in the plot for $\theta = 6^{\circ}$, where two clear peaks 
are produced---one arising from the stream overflow and one 
from the disk structure. Since the disk precesses in the frame
of the binary (for low $q$ systems), the absorption feature arising solely
from the eccentric disk structure should vary linearly
in phase with time. Conversely, structures generated directly by the
stream-disk interaction should remain at fixed phases 
(regardless of the mass ratio of the binary).

\subsection{Fate of Stream Material}

\begin{figure}[t]
 \hspace{0.9truein}
 \psfig{figure=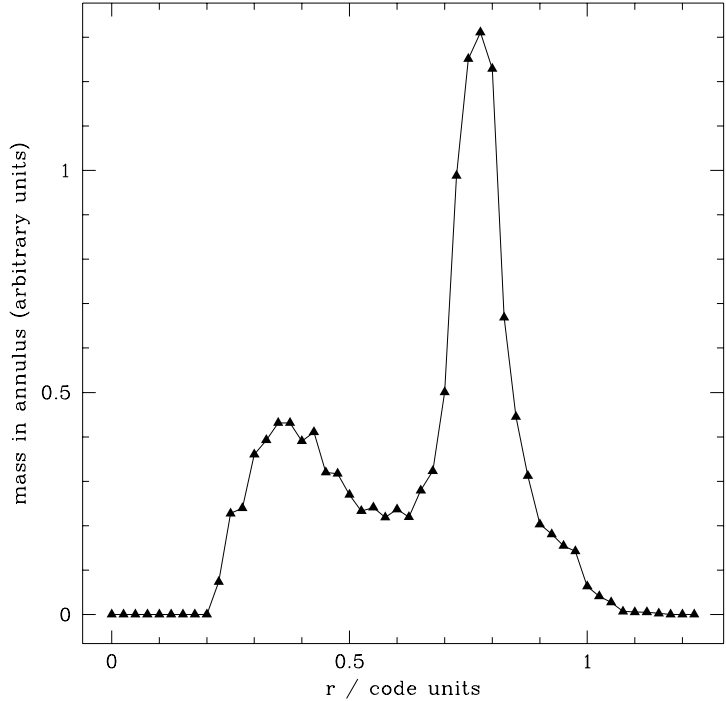,width=4.0truein,height=4.0truein}
 \caption{Destination of stream mass (in arbitrary units) as a
function of radius.} 
 \label{fig7}
\end{figure}

Time-dependent models for accretion disks in (dwarf nova) systems vulnerable to
thermal instability-driven outbursts generally add mass to the
disk from the stream either in the outermost computational
zone (\eg Pringle, Verbunt \& Wade 1986), or else with a
narrow gaussian profile at the outer edge (\eg Cannizzo 1993).
Neither is likely to be a good representation of the 
physical situation if a significant
fraction of the stream mass overflows the disk rim and circularises
at smaller radii. To examine this possibility, Fig.~7 shows the
destination of stream material as a function of radius in the
disk, for simulation LMXB1, where the disk is circular and
the complicating effects of an eccentric disk are not
present. The distribution is bimodal, with most of the mass
being deposited and circularising at the outer disk rim,
but with a significant fraction ($\sim 1/3$) ending up
at much smaller radii. The outer peak is caused by 
the dense inner core of the stream, which on impact with
the disk becomes buried in the outer edge, while the inner
peak (near the stream circularisation radius) is formed
from the stream gas that overflows the disk rim.

This result must be treated with some caution however, as the current
simulations do not have sufficient resolution to resolve,
for example, any surface instabilities that might entrain the
overflowing stream gas more efficiently and thus keep it
at larger radii. Calculations with a higher resolution and
better treatment of the thermal properties of the gas will
be required to investigate this possibility. The general 
picture from our simulations, however, is that neither mass
addition exclusively at the outer edge nor mass addition 
primarily at the circularization radius is likely to prove
a good approximation. Rather, the surface density increment
is bimodal with roughly equal amplitudes at the two radii, and we
anticipate that this
is likely to alter the detailed behaviour seen in time-dependent
disk instability models.

\subsection{Location of energy dissipation}

In our simulations artificial viscosity is used in the
momentum equation to generate shocks and prevent particles
streaming freely through each other. In the physical disk,
these shocks would lead to strong heating of the gas, and
create bright spots on the disk surface. The strength of
this dissipation can be estimated by calculating the 
shock heating that would occur in the simulation if the
gas had an adiabatic ($\gamma=5/3$) equation of state,
while the flow pattern remained as we have described. 

\begin{figure}[t]
 \hspace{0.5truein}
 \psfig{figure=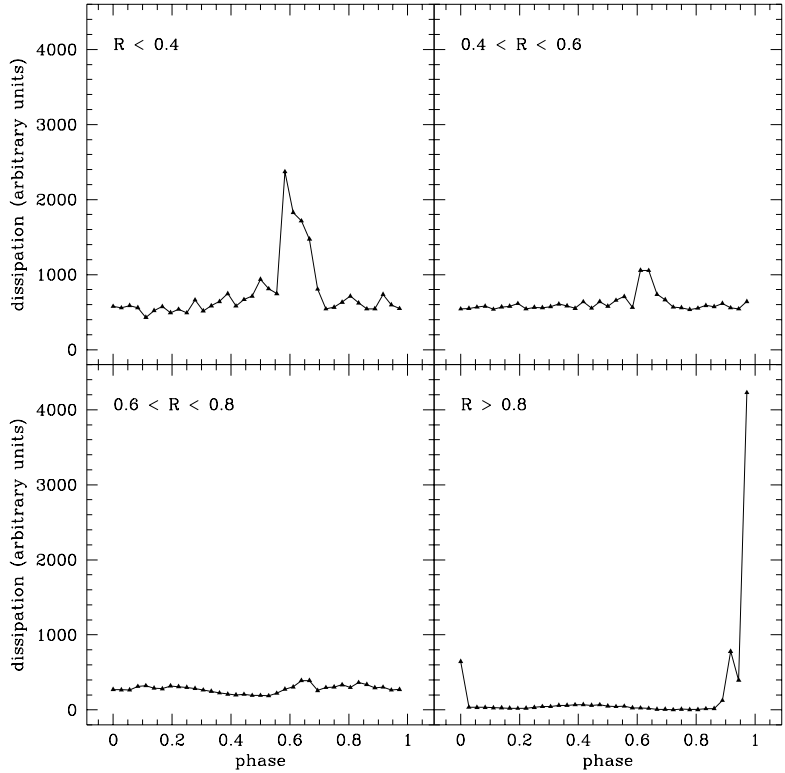,width=4.5truein,height=4.5truein}
 \caption{Rate of heating from shocks as a function of phase
for annuli in the disk (simulation LMXB1). The panels run
from the innermost annulus at upper left, to the outermost at
bottom right. The disk edge is at $R \approx 0.9$ in these
units.} 
 \label{fig8}
\end{figure}

Fig.~8 shows the energy dissipation estimated in this
way as a function of phase, for annuli that cover
the radial extent of the disk. For the annuli at smaller
radii, the background level of dissipation is uniformly greater 
because of the increased viscous dissipation that is unconnected
with shocks. Excesses over this level represent areas where
bright spots might be expected to occur. As expected, the strongest
dissipation is found in the outermost annulus, at the
location where the stream strikes the disk. However there
is also strong shock heating occuring in the innermost ring,
at the location where the overflowing stream gas collides
again with the disk (at phase $\phi \sim 0.6$). 
A smaller fraction of stream material
reaches this radius, but being being deeper in
the potential well the total shock heating is similar.
No prominent features are seen in the two intermediate 
annuli.

The width of the features seen in Fig.~8 represents
only the azimuthal extent of the regions where shock heating 
would occur, the actual size of the
bright spots produced as the shocked gas cools will undoubtably 
be much larger and may cover a sizeable fraction of the
disk circumference (Rozcyczka \& Schwarzenberg-Czerny 1987).
Our calculations suggest that a second bright spot
should be formed at the roughly diametrically opposed
phase where the overflowing gas returns to the disk
plane. This inner bright spot is of comparable luminosity
to that formed directly by the stream-disk impact, though
relative to the background emission at the smaller radius it is,
of course, much weaker. It would thus be most easily seen
in systems with low viscosity (\eg dwarf novae with long
recurrence times), and could provide valuable clues to
the degree of stream overflow occuring in these systems.

\section{SUMMARY AND DISCUSSION}

We have presented three dimensional
SPH calculations of the stream-disk interaction.
The main conclusion to be drawn from these simulations
is that significant amounts of stream material are
able to overflow or bounce off the edge of the disk,
and flow over the disk surface towards smaller radii.
This creates a peak in the column density towards the
primary at a phase of around 0.8, and is a likely
cause of the absorption dips observed at around this
phase in some nearly edge-on low-mass X-ray binaries (see \eg review
by White, Nagase, \& Parmar 1995).
If this is the case, the absorbing material is located (in radial
distance) intermediate between the disk edge and the stream
circularization radius. At the highest angles of
elevation above the disk surface ($\theta \age 10^{\circ}$),
we see no significant absorbing column at any other phases.
Recent EUVE observations of U~Gem in outburst (Long
\etal 1996) show an absorption dip (at phase $\sim 0.75$) caused
by material a few scale heights above the disk surface, in
complete agreement with the results of the present study.  ASCA
observations show that the dips persist in quiescence (Szkody
\etal 1996).

This result differs from that of Hirose, Osaki \& Minishige
(1991), who in addition to predicting absorption at phase
0.8 also found absorption at phase 0.2 in their models
(although as they did not compute column densities from
their simplified calculation it is unclear whether both those
features would be visible at large $\theta$). We do
find significant absorption at around phase 0.1--0.2 for
lower angles of elevation above the disk plane, and
this seems to be primarily due to the eccentricity induced
in the disk by the gravity of the companion. We note that
in low mass ratio systems absorption features generated
by the eccentric disk structure should precess with the
disk, and thus can be separated from those induced
by the stream-disk interaction, which remain at a fixed phase
($\sim 0.8$) regardless of the mass ratio.  Therefore, in systems
where two absorption features are seen, a straightforward
observational test to determine their origin is possible, by
monitoring for a prolonged period the phases of maximum
absorption.

The presence of gas that is not in hydrostatic equilibrium
flowing over the disk raises a number of intriguing possibilities.
If exposed to an X-ray power-law spectrum from the central
source, it may be prone to a two-phase instability and
break up into dense clouds surrounded by a hot diffuse
medium (Krolik, McKee \& Tarter 1981; Frank, King \& Lasota 1987). 
Such a picture
might explain the very short-timescale variations seen
in some dipping X-ray sources. Less speculatively, the
wide range of radii over which mass from the stream
circularises in our simulations, implies that significant
changes might be expected in the behavior predicted from
time-dependent thermal disk instability models. In particular,
during quiescence, the quasi ballistic flow over the disk might
dominate inflow caused by viscosity in the outer disk, and
this would affect predictions for the behaviour of
the disk radius. Significant stream overflow implies that
less material with low specific angular momentum will
be added at the outer edge of the disk, and hence less
contraction of the disk radius during quiescence would be
expected. If the degree of overflow were large (as might occur
if the outer disk were cold relative to the mass-donating
star), then substansive differences in the surface density 
profile during quiescence would also occur. This could lead,
for example, to thermal instabilities triggered at intermediate
disk radii, and other, detailed, differences to standard models.

For systems that undergo outbursts, the presence of overflowing
stream gas provides a possible observational probe of the
disk radius. As the disk expands, the phase of maximum
absorption is expected to increase (closer to the phase of
eclipse) as a consequence both of the position and angle at
which the accretion stream reaches the disk. More subtle
effects might also occur due to the varying temperature of
the outer disk over the course of an outburst cycle. A
hotter disk is thicker, and this should reduce the
amount of stream material able to overflow the rim and thereby
reduce the depth of the absorption. A thicker disk might
of course absorb radiation from the primary
directly, but this effect would be axisymmetric and thus
distinguishable from that caused by high $|z|$ stream
gas. Simultaneously {\em more}
energy would be dissipated in the bright spot at the
disk rim, so that the bright spot luminosity would increase
as the outburst commenced. There is some observational
evidence that this brightening does occur (Livio 1994).

We should note that in their attempt to model the visual light
curve of the eclipsing supersoft X-ray source CAL~87, Schandl,
Meyer-Hofmeister, \& Meyer (1996), found that they had to assume
the presence of an optically thick ``spray'' of material around
the disk.  They conjectured that such a spray might form by the
disk-stream interaction.  An examination of Fig.~2 reveals that
we indeed obtain a configuration similar to that assumed by
Schandl \etal (1996).

Finally we remark that in this paper we have concentrated 
on explaining features
seen in the X-ray and EUV wavebands. Important information
is also available in the optical, especially from the
analysis of time-resolved spectra using doppler tomography
(Marsh \& Horne 1988). Many systems have now been mapped
using this technique (\eg Kaitchuck \etal 1994), and in some
cases the path of the accretion stream can be clearly traced.
A future comparison of synthetic doppler maps generated from simulations
of the type presented here with these data appear to provide a fruitful
direction in which the research into stream-disk
interactions can continue.

\subsection*{ACKNOWLEDGEMENTS}

We thank Ian Bonnell, Cathie Clarke, Melvyn Davies, Andy Fabian,
Steve Lubow and Donald Lynden-Bell for useful discussions
and suggestions. PJA thanks Space
Telescope Science Institute for hospitality.  ML acknowledges
support from NASA Grant NAGW-2678 at ST~ScI, and thanks the
Institute of Astronomy, Cambridge, for its hospitality.

\end{document}